\documentclass[twocolumn,aps,showpacs,floatfix,prc]{revtex4}
\usepackage[dvips]{epsfig}
\begin{document}

\title{  $J/\psi$  production  in  Au+Au/Cu+Cu collisions at $\sqrt{s}_{NN}$=200 GeV and
the threshold model}

\author{A. K. Chaudhuri}
\email[E-mail:]{akc@veccal.ernet.in}
\affiliation{Variable Energy Cyclotron Centre,\\ 1/AF, Bidhan Nagar,
Kolkata 700~064, India}

\begin{abstract}
Using the QGP motivated threshold model, where all the $J/\psi$'s
are  suppressed  above  a threshold density, we have analyzed the
preliminary PHENIX data on the centrality  dependence  of 
nuclear modification factor for $J/\psi$'s in Cu+Cu and in Au+Au collisions, at RHIC energy, $\sqrt{s}_{NN}$=200 GeV. Centrality dependence of  $J/\psi$ suppression in Au+Au collisions are well explained in the model for threshold densities in ranges of 3.6-3.7 $fm^{-2}$. $J/\psi$ suppression in Cu+Cu collisions on the other hand are not explained in the model.  
 \end{abstract}  
\maketitle
 
\section{introduction} 
In  relativistic  heavy  ion  collisions $J/\psi$ suppression has
been recognized as an important tool  to  identify  the  possible
phase transition to quark-gluon plasma. Because of the large mass
of  the  charm  quarks,  $c\bar{c}$ pairs are produced on a short
time scale. Their tight binding also makes them immune  to  final
state interactions. Their evolution probes the state of matter in
the  early  stage  of the collisions. Matsui and Satz  \cite{Matsui:1986dk}, 
predicted that in presence of quark-gluon plasma  (QGP),  binding
of a $c\bar{c}$  pair  into  a  $J/\psi$  meson will be hindered,
leading to the  so  called  $J/\psi$  suppression  in  heavy  ion
collisions  \cite{Matsui:1986dk} .  Over  the  years,  several  groups have
measured the $J/\psi$ yield in heavy ion collisions (for a review
of the data prior to RHIC energy collisions, and the interpretations see Refs.  \cite{vo99,ge99}).
In  brief,  experimental  data do show suppression. However, this
could be attributed to the conventional nuclear absorption,  also
present in $pA$ collisions.

In  recent  Au+Au  collisions  at  RHIC,  one  observe a dramatic
suppression of hadrons with high  momentum,  transverse  to  beam
direction            (high           $p_T$           suppression)
\cite{BRAHMSwhitepaper,PHOBOSwhitepaper,PHENIXwhitepaper,STARwhitepaper}.
This has been interpreted as evidence for the  creation  of  high
density,  color  opaque  medium  of  deconfined quarks and gluons
\cite{QGP3jetqu}. It is expected that high density, color opaque medium
will leave its imprint on $J/\psi$ production. At RHIC energy, it has been
argued that rather than suppression, charmoniums will be enhanced
\cite{ Thews:2000rj,Braun-Munzinger:2000px}. 
Due to large initial energy, large number of $c\bar{c}$ pairs will be
produced in initial hard scatterings. Recombination of $c\bar{c}$
can occur enhancing the charmonium production.
PHENIX  collaboration  have measured the centrality dependence of
$J/\psi$ invariant yield in  Au+Au  collisions  at  RHIC  energy,
$\sqrt{s}_{NN}$=200  GeV \cite{Adler:2003rc,Nagle:2002ib}.  More recently they have
improved  upon  the  statistics  and   preliminary   results   for the
centrality  dependence  of nuclear modification factor $(R_{AA})$
for $J/\psi$ in Au+Au collisions  and  in  Cu+Cu  collisions  are
available  \cite{Nagle:2005sh,PereiraDaCosta:2005xz}. In most central Au+Au collisions, $J/\psi$'s are suppressed by a factor of $\sim 3$.
PHENIX data on
$J/\psi$ production in Au+Au/Cu+Cu collisions, are not consistent
with models which predict $J/\psi$ enhancement
\cite{ Thews:2000rj,Braun-Munzinger:2000px}.  It was also  seen
that  various  models,  e.g.  comover  model   
\cite{Capella:2000zp}, statistical  coalescence  model    
\cite{Kostyuk:2003kt}  or the kinetic model  \cite{Gorenstein:2000ck},
fail to explain the   (preliminary) PHENIX data  on  the  nuclear
modification   factor   for   $J/\psi$  in  Cu+Cu  and  in  Au+Au
collisions. The data are also not  explained  in  the Glauber model of normal  nuclear
absorption \cite{Vogt:2005ia}. Recently, in a QCD based 
nuclear absorption model , we have analyzed
the preliminary PHENIX data on $J/\psi$ suppression in Cu+Cu
and in Au+Au collisions \cite{Chaudhuri:2003sv}. In the 
QCD based nuclear absorption model 
\cite{Qiu:1998rz,Chaudhuri:2001zx}, $c\bar{c}$ pair, during its passage through a nuclear medium,
gain relative 4-square momentum. Some of the pairs gain enough to cross the open charm threshold and are lost.  The model  explained the PHENIX data on the centrality dependence of $J/\psi$ suppression in Cu+Cu collisions at RHIC but failed for
Au+Au collisions. It was concluded that in Au+Au collisions, $J/\psi$ are suppressed in a medium, unlike that produced in SPS energy nuclear collisions or at RHIC energy Cu+Cu collisions. If in Au+Au collisions, $J/\psi$'s are
suppressed in a deconfined matter, PHENIX data should be explained in a QGP motivated model, like the threshold model  \cite{Blaizot:2000ev,Blaizot:1996nq}.
Blaizot  et  al  \cite{Blaizot:2000ev,Blaizot:1996nq},  proposed the threshold
model  to  explain  the NA50 data on  anomalous $J/\psi$ suppression in 
158 AGeV Pb+Pb 
collisions at SPS energy \cite{Abreu:2000ni} .  
To mimic the onset of deconfining phase transition
above  a  critical  energy  density  and  subsequent  melting  of
$J/\psi$'s, $J/\psi$ suppression was linked with the local energy
density.  If  the  energy  density at the point where $J/\psi$ is
formed, exceeds a critical  value  ($\varepsilon_c$),  $J/\psi$'s
disappear.   

In the present paper, in the threshold model, we have analyzed the preliminary PHENIX data on the centrality dependence of $J/\psi$ suppression in Cu+Cu and in Au+Au collisions. As it will be shown below, while the centrality dependence of $J/\psi$ suppression in Au+Au collisions are well explained in the model, that in Cu+Cu collisions are not. We have also analyzed the PHENIX data on the centrality dependence of
$p_T$ broadening of $J/\psi$ \cite{Nagle:2005sh,PereiraDaCosta:2005xz}. No definitive conclusions can be obtained from the $p_T$ broadening  data .

The plan of the paper is as follows: in section II, we briefly describe the threshold model. PHENIX data on the centrality
dependence of $J/\psi$ suppression are analyzed in section III.
In section IV, we analyze the PHENIX data on $p_T$ broadening
of $J/\psi$. Summary and conclusions are drawn in section V.

\section{Threshold model}

The details of the threshold model could be found in  \cite{Blaizot:2000ev,Blaizot:1996nq}.
It is assumed that fate of a $J/\psi$ depend on the local energy
density, which is proportional to participant density. If the energy density or equivalently, the participant density, exceeds a critical or threshold value, deconfined matter is formed and all the $J/\psi$'s are completely destroyed (anomalous suppression). 
This anomalous suppression is in addition to the
"conventional nuclear absorption".  
Transverse expansion of the system is neglected. It is implicitly assumed that $J/\psi$'s are absorped
before the transverse expansion sets in.

In the threshold model, number of $J/\psi$ mesons, produced in a AA collision, at impact parameter ${\bf b}$ can be written as,

\begin{eqnarray} \label{eq1}
\sigma^{J/\psi}_{AA}({\bf b}) = &&\sigma^{J/\psi}_{NN} \int  
 d^2{\bf s}  
 T^{eff}_A({\bf s}) T^{eff}_B({\bf b-s}) \nonumber \\
  &&\times S_{anom}({\bf b,s}),
\end{eqnarray}
  
\noindent  where $T^{eff}(b)$ is the effective nuclear thickness,

\begin{equation} \label{eq2}
T^{eff}({\bf  b})=\int_{-\infty}^{\infty}  dz  \rho({\bf   b},z)
exp(-\sigma_{abs}  \int_z^{\infty} dz\prime \rho({\bf
b},z\prime)),
\end{equation}

\noindent $\sigma_{abs}$ being the $J/\psi$-Nucleon absorption cross-section.  
$S_{anom}({\bf   b,s})$  in  Eq.\ref{eq1}  is  the  anomalous
suppression factor introduced by Blaizot {\em et al.}  \cite{Blaizot:2000ev,Blaizot:1996nq}. Assuming that  all  the
$J/\psi$'s  get suppressed above a threshold density ($n_c$), the
anomalous suppression can be written as,

\begin{equation}  \label{eq3}
S_{anom}({\bf b,s}) =\Theta (n({\bf b,s})-n_c) 
\end{equation}

\noindent   where  $n_c$ is the critical or the threshold density.
$n({\bf b,s})$   is  the  local transverse  density.
At impact parameter ${\bf b}$ and at the transverse position ${\bf s}$, local transverse density it can be obtained as,

\begin{eqnarray}
n({\bf b,s})=&&T_A({\bf s})[1-exp(-\sigma_{NN} T_B({\bf s}-{\bf b}))] \nonumber \\
&&+T_B({\bf b}-{\bf s})[1-exp(-\sigma_{NN} T_A({\bf s}))]
\end{eqnarray}

Blaizot {\em et al} \cite{Blaizot:2000ev,Blaizot:1996nq} fitted the NA50 data \cite{Abreu:2000ni} on
the transverse energy dependence of $J/\psi$ suppression in 158 AGeV
Pb+Pb collisions and obtain the threshold density $n_c$.  
With $J/\psi$-nucleon absorption cross-section $\sigma_{J/\psi N}$=6.4 mb, NA50 data  are well explained in the model with   $n_c$=3.7 $fm^{-2}$. Better fit to the data is obtained if the
theta function (Eq.\ref{eq3}) is smeared, at the expense of an additional parameter. Later experiments \cite{Cortese:2003iz} indicate that $J/\psi$-nucleon absorption cross-section is $\sim$ 4 mb, rather than 6.4 mb.   NA50 collaboration also revised their data \cite{na50a}. The revised NA50 data were also analyzed in the
threshold model  \cite{Chaudhuri:2003nj}. With 
$\sigma_{abs} \sim$ 4 mb,   large smearing of the threshold density is required. Large smearing of threshold density, effectively excludes formation of deconfined matter at SPS 
energy.

\begin{figure}[h]
\centerline{\psfig{figure=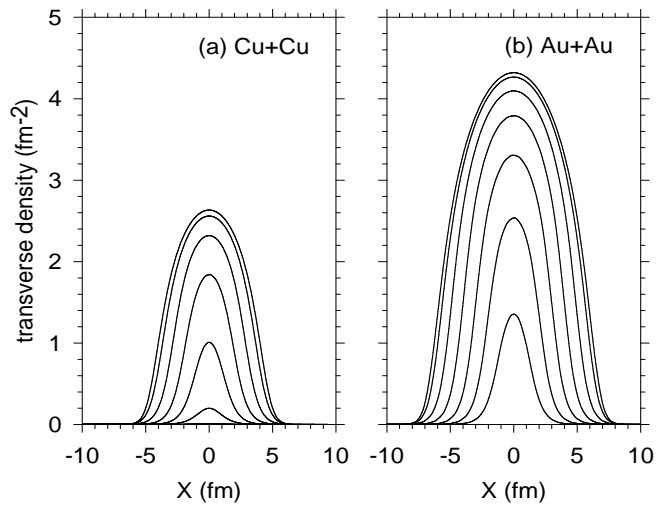,height=12cm,width=9cm}}
\vspace{-4.5cm}   \caption{Transverse density in Cu+Cu (left panel) and in Au+Au (right panel) collisions, for various values of the impact parameter,
b=0,2,4.... (from top to bottom). The origin is at a distance,
$d=b/(1+R_A/R_B)$ from the center of the nucleus A.}
\label{F1}
\end{figure}

\section{$J/\psi$ suppression in Cu+Cu/Au+Au collisions}

In the threshold model, fate of a $J/\psi$ is determined on the
local (transverse) density. If the local (transverse) density exceeds
the threshold density, $J/\psi$'s are completely destroyed.  
In Fig.\ref{F1}, for a number of impact parameters, the  
transverse density, $n({\bf b,s})$ in Cu+Cu and  in Au+Au collisions are shown. We have used   
the Woods-Saxon form for the density, 

\begin{equation}
\rho(r)=\frac{\rho_0}{1+exp((r-R)/a)}, \hspace{1cm}\int d^3r \rho(r)=A
\end{equation}
 
\noindent with $R$=4.456 (5.415) fm and $a$=0.54 (0.535) fm,
for Cu (Au) nuclei. 

For central collisions, maximum transverse
density in Cu+Cu collisions is $ \sim$2.63 $fm^{-2}$, while that for
Au+Au collisions is $\sim$ 4.32 $fm^{-2}$. 
Then if $J/\psi$'s are anomalously suppressed,
 say, above a threshold density, $n_c$=3.7 $fm^{-2}$, 
$J/\psi$ suppression
in Cu+Cu collisions will not be affected as the transverse density never exceeds the threshold density. In
Au+Au collisions, on the other hand, $J/\psi$'s will be anomalously suppressed. In Au+Au collisions also,   only in collisions
where local density $n({\bf b,s})$ exceeds the threshold density,
$J/\psi$'s will be anomalously suppressed. In all other collisions,   $J/\psi$'s will be  absorped only due to $J/\psi$-nucleon interaction. Then if $J/\psi$ suppression is measured as a function of impact parameter or equivalently,
as a function of centrality of collisions, sudden change of slope will be observed.
 
\begin{figure}[h]
\centerline{\psfig{figure=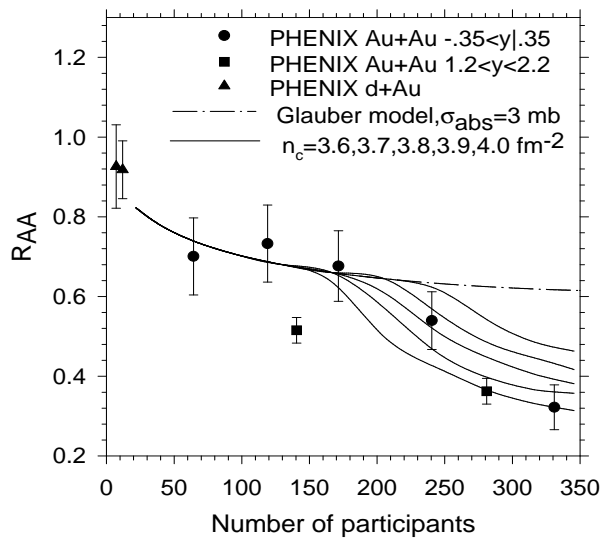,height=12cm,width=9cm}}
\vspace{-4.5cm}   \caption{Preliminary PHENIX data on the participant number dependence of nuclear modification factor ($R_{AA}$)
for $J/\psi$, in Au+Au collisions. $R_{AA}$ for $J/\psi$ in d+Au
collisions are also shown. The dashed line is the Glauber model
 prediction for $R_{AA}$ with $\sigma_{abs}$=3 mb.  The solid
lines (bottom to top) are the threshold model predictions for threshold density,
$n_c$=3.6,3.7,3.8,3.9 and 4 $fm^{-2}$.}
\label{F2}
\end{figure}

Such a sudden change of slope in $J/\psi$ suppression is
observed by the PHENIX collaboration in Au+Au collisions.
In   Fig.\ref{F2},   preliminary  PHENIX  data  on  the
centrality dependence of nuclear modification  factor  ($R_{AA}$)
for  $J/\psi$,  in Au+Au collisions are shown
\cite{Nagle:2005sh,PereiraDaCosta:2005xz}.
PHENIX collaboration has taken data   in two 
ranges of rapidity intervals, (i) $-0.35 \leq y \leq  0.35$  and 
(ii)$1.2  \leq  y  \leq  2.2$. Both the data are shown.
Even though data points are few, sudden change of slope of $R_{AA}$ around $N_{part} \sim 150$ is evident. 
PHENIX data for $R_{AA}$ in d+Au collisions are also shown in the Figure. $J/\psi$'s are suppressed in d+Au collisions also.
Glauber model analysis indicate that  in d+Au collisions,
 $J/\psi$-nucleon absorption cross-section is small, $\sigma_{abs} \approx 1-3$ mb \cite{Vogt:2005ia}.  
In Fig.\ref{F2},  
prediction for $R_{AA}$, in the   Glauber model of normal nuclear
absorption is  shown as the
dash-dot-dashed line. The prediction is obtained with $\sigma_{abs}$=3 mb.   It is interesting to note that 
the Glauber model of nuclear absorption, with $\sigma_{abs}$=3 mb,  explains the suppression in peripheral and mid-central collisions.
Only in very central collisions, the Glauber model of nuclear absorption model predict much less
suppression than observed by the PHENIX collaboration.
Anomalous suppression in Blaizot's threshold model can provide
the additional suppression required in very central collisions.
Due to paucity of data, we do not attempt to fit the PHENIX data  
and extract the threshold density, $n_c$. Rather, for a number  of threshold density, $n_c$=3.6,3.7,3.8,3.9 and 4.0 $fm^{-2}$,  we obtain
predictions for $R_{AA}$ in the threshold model.  
In Fig.\ref{F2}, solid lines are  the predictions in the threshold
model. With anomalous suppression, $J/\psi$'s are strongly suppressed in central collisions and it is evident that for 
threshold density in the ranges of
3.6-3.7 $fm^{-2}$, the threshold model describe the data adequately well.

\begin{figure}[h]
\centerline{\psfig{figure=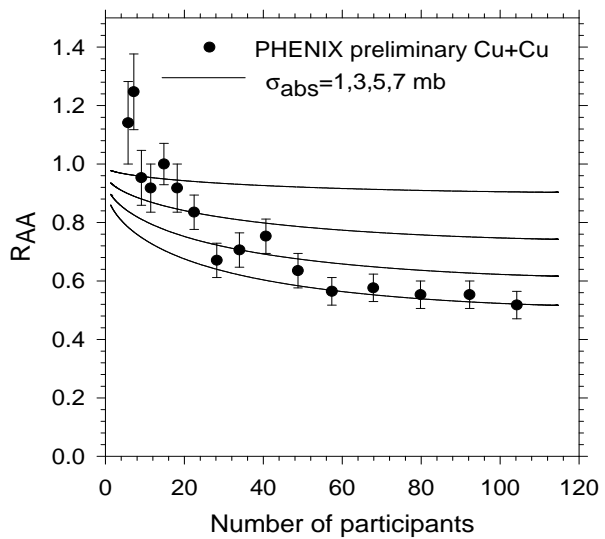,height=12cm,width=9cm}}
\vspace{-4.5cm}   \caption{Preliminary PHENIX data on the centrality dependence of nuclear modification factor ($R_{AA}$)
for $J/\psi$ in Cu+Cu collisions. $R_{AA}$ for $J/\psi$ in d+Au
collisions are also shown. The solid
lines (top to bottom) are Glauber model predictions with $\sigma_{abs}$=
1,3,5 and 7 mb.}
\label{F3}
\end{figure}
We now analyze the PHENIX data on the centrality dependence 
of nuclear modification factor, in Cu+Cu collisions. The data are 
shown in Fig.\ref{F3}. We have not distinguished between 
the mid-rapidity and forward rapidity data. In contrast to Au+Au data, $R_{AA}$ in Cu+Cu collisions do not show any sudden change in slope. Threshold model, which induces sudden change
in slope, is not warranted by the data. 
Indeed, if the threshold density is in the ranges of 3.6-3.7 $fm^{-2}$,
anomalous suppression will be ineffective in
Cu+Cu collisions. As shown in Fig.\ref{F1}, maximum  transverse density
reached in central Cu+Cu collisions is $\sim$ 2.63 $fm^{-2}$, much less than the threshold density 3.6-3.7 $fm^{-2}$. 

As mentioned earlier, $J/\psi$
production in d+Au collisions at RHIC, require $\sigma_{abs} \sim$ 1-3 mb. Glauber model predictions with $\sigma_{abs}$=3 mb, also explains the
PHENIX data on the centrality dependence of $J/\psi$ suppression in peripheral and mid-central  Au+Au collisions. Only in very central Au+Au collisions,  data demand anomalous suppression. 
However, in Cu+Cu collisions, Glauber model predictions with $\sigma_{abs}$=1-3 mb fails to explain the centrality dependence of $J/\psi$ suppression. 
In Fig.\ref{F3}, Glauber model predictions for $J/\psi$ suppression
in Cu+Cu collisions,
for 
various values of $J/\psi$-nucleon absorption cross-sections, $\sigma_{abs}$=1,3,5, and 7 mb are shown. For $\sigma_{abs}$=1-3, the model could explain suppression only
in very peripheral collisions ($N_{part}$=10-20). In mid-central and very central collisions, the model with $\sigma_{abs}$=1-3 mb, underpredict the
suppression.  If the $J/\psi$-nucleon cross section is increased, while the model explains the mid-central and very central collisions, it fails to explain peripheral collisions. The analysis indicate that, with a single value $J/\psi$-nucleon absorption cross-section $\sigma_{abs}$, the Glauber model of nuclear absorption,  can not explain the centrality dependence of $J/\psi$ suppression in Cu+Cu collisions.   $J/\psi$ suppression in Cu+Cu collisions at RHIC energy is more complex than envisaged in the Glauber model of nuclear absorption. Indeed, a more complex,
QCD based nuclear absorption model, does explain the centrality dependence of $J/\psi$ suppression in Cu+Cu collisions adequately well \cite{Chaudhuri:2003sv}.

\begin{figure}[h]
\centerline{\psfig{figure=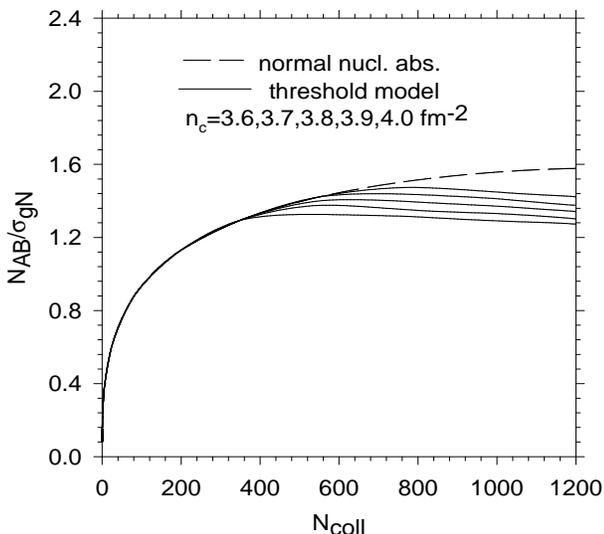,height=12cm,width=9cm}}
\vspace{-4.5cm}   \caption{Collision number dependence of the
ratio $N_{AB}/\sigma_{gN}$ in Au+Au collisions. The dashed line is the ratio in the normal nuclear absorption model, with $\sigma_{abs}$=3 mb. The solid lines
are $N_{AB}/\sigma_{gN}$ in the threshold
model, with threshold density $n_c$=3.6,3.7,3.8,3.9 and 4.0
$fm^{-2}$ (from bottom to top) respectively.}
\label{F4}
\end{figure}

In the threshold model, it is implicitly assumed that $J/\psi$'s
are absorped in a deconfined matter. The critical energy density for deconfined matter formation is  proportional to the threshold density. Melting of $J/\psi$ due to color screening is mimicked by the sudden on set of suppression. 
Successful description of PHENIX data on $J/\psi$ suppression in Au+Au collisions then strongly support formation of  deconfined matter in central Au+Au collisions. However, we note that the model neglects some very important effects, e.g. 
(i) feed back
from $\psi^\prime$ and $\chi$ states and (ii) transverse expansion.
A considerable fractions of $J/\psi$'s are  from decay of $\psi^\prime$ and $\chi$ states. That part is completely neglected here. Threshold density for  anomalous suppression of higher states, $\psi^\prime$ and $\chi$ should be less than that for a $J/\psi$. Then presently estimated threshold density $n_c$=3.6-3.7 $fm^{-2}$ represent an upper limit of the threshold density.
At RHIC, model studies indicate that in the deconfined phase, the system undergoes significant transverse expansion  \cite{Kolb:2003dz}.
The local transverse density is a key ingredient  to the Threshold model. In an expanding system, local transverse density will be diluted.  
$J/\psi$'s which   are anomalously suppressed in a static system, may survive in an expanding system due to dilution. Then,  the
presently estimated critical density 3.6-3.7 $fm^{-2}$ will
again
represent an upper limit of the threshold density. 

\section{centrality dependence of $p_T$ broadening in Cu+Cu/Au+Au collisions}

 It  is  well  known  that  in pA and AA collisions, the secondary
hadrons     generally     show            $p_T$      broadening
\cite{ptbr,Kharzeev:1997ry}. $p_T$ broadening of $J/\psi$ in
Cu+Cu and in Au+Au collisions at RHIC energy $\sqrt{s}_{NN}$=200 GeV, has been measured by the 
PHENIX collaboration \cite{Nagle:2005sh,PereiraDaCosta:2005xz}.
They measured the collision number  dependence of square of transverse momentum for $J/\psi$. 
It is interesting to compare the threshold model predictions with the PHENIX data.

The  natural  basis for the $p_T$ broadening is the initial state
parton  scatterings.  For  $J/\psi$'s,  gluon  fusion  being  the
dominant  mechanism  for  $c\bar{c}$  production,  initial  state
scattering   of   the   projectile/target   gluons    with    the
target/projectile   nucleons   causes   the   intrinsic  momentum
broadening of  the  gluons,  which  is  reflected  in  the  $p_T$
distribution  of  the  resulting  $J/\psi$'s.  Parameterising the
intrinsic transverse momentum of a gluon, inside a nucleon as,

\begin{equation} f(q_T) \sim exp(-q^2_T/<q^2_T>) \end{equation}

\noindent  momentum  distribution of the resulting $J/\psi$ in NN
collision is obtained by convoluting two such distributions,

\begin{equation}             f^{J/\psi}_{NN}(p_T)            \sim
exp(-p^2_T/<p^2_T>^{J/\psi}_{NN}) \end{equation}

\noindent  where  $<p^2_T>^{J/\psi}_{NN}  =  <q^2_T>+<q^2_T>$. In
nucleus-nucleus collisions at  impact  parameter  ${\bf  b}$,  if
before  fusion, a gluon undergo random walk and suffer $N$ number
of subcollisions, its square momentum  will  increase  to  $q^2_T
\rightarrow  q^2_T  +  N\delta_0$,  $\delta_0$  being the average
broadening in each subcollisions.  Square  momentum  of  $J/\psi$
then easily obtained as,

\begin{equation}     \label{eq11}     <p^2_T>^{J/\psi}_{AB}(b)     =
<p^2_T>^{J/\psi}_{NN} + \delta_0 N_{AB}({\bf b}) \end{equation}

\noindent  where $N_{AB}({\bf b})$ is the number of subcollisions
suffered by the projectile and target gluons with the target  and
projectile nucleons respectively.

Average number of collisions $N_{AB}({\bf b})$ can be obtained in
a Glauber model \cite{Kharzeev:1997ry}. At impact parameter ${\bf
b}$,  the  positions  $({\bf  s},z)$  and  $({\bf b-s},z^\prime)$
specifies the formation point of $c\bar{c}$ in  the  two  nuclei,
with ${\bf s}$ in the transverse plane and $z,z^\prime$ along the
beam  axis.  The  number  of collisions, prior to $c\bar{c}$ pair
formation, can be written as,

\begin{eqnarray}  \label{2}  N(b,s,z,z^\prime)  =  && \sigma_{gN}
\int_{-\infty}^z dz_A \rho_A(s,z_A) \\ \nonumber && + \sigma_{gN}
\int_{-\infty}^{z^\prime}        dz_B        \rho_B(b-s,z^\prime)
\end{eqnarray}

\noindent where $\sigma_{gN}$ is the gluon-nucleon cross-section.
Above  expression  should  be  averaged  over  all  positions  of
$c\bar{c}$ formation with  a  weight  given  by  the  product  of
nuclear densities and survival probabilities $S$,

\begin{eqnarray}\label{3}     &&N_{AB}(b)=     \int     d^2     s
\int^\infty_{-\infty}   dz   \rho_A(s,z)    \int^\infty_{-\infty}
dz^\prime    \rho_B(b-s,z^\prime)    \times    \nonumber  \\
 &&
S(b,s,z,z^\prime)     N(b,s,z,z^\prime)     /      \int      d^2s
\int^\infty_{-\infty}  dz  \rho_A(s,z)  \times    \nonumber \\ &&
\int^\infty_{-\infty}       dz^\prime        \rho_B(b-s,z^\prime)
S(b,s,z,z^\prime) \end{eqnarray}

Centrality dependence of the ratio $N_{AB}/\sigma_{gN}$, in Au+Au collisions,  for the threshold densities,
$n_c$=3.6,3.7,3.8,3.9 and 4.0 $fm^{-2}$, are shown in Fig.\ref{F4}, (the solid lines from bottom to top). We also show the ratio in the normal nuclear absorption model (the dashed line).
$N_{AB}/\sigma_{gN}$  increases with centrality, more central the collisions, the gluons suffer more number of collisions. In
normal nuclear absorption model, $N_{AB}/\sigma_{gn}$ continues to increase with centrality (or collision number). However, rate of increase slows down at more central collisions.
A different behavior is obtained in the threshold model. For a fixed threshold density $n_c$, $N_{AB}/\sigma_{gn}$ exactly corresponds to normal nuclear absorption model, till 
a collision number $N_c$. Beyond $N_c$,
$N_{AB}/\sigma_{gN}$ hardly changes. It is understood. 
Beyond a $N_c$, transverse density exceeds the threshold density and $J/\psi$'s are completely destroyed. As $N_{AB}$ is weighted by the anomalous suppression, it hardly changes beyond that collision number.
 
\begin{figure}[h]
\centerline{\psfig{figure=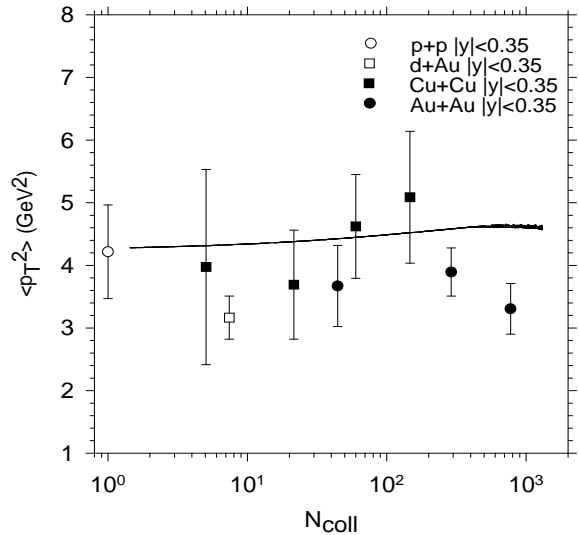,height=12cm,width=9cm}}
\vspace{-4.5cm}   \caption{$J/\psi$ mean square transverse momentum as a function of collision number, in 
mid-rapidity p+p, d+Au, Cu+Cu and Au+Au collisions are shown. The solid line  is the best fit,  in the
threshold model, to the combined
Cu+Cu and Au+Au data.}
\label{F5}
\end{figure}

$p_T$  broadening  of  $J/\psi$'s in AA collisions depends on two
parameters,  (i)  $<p^2_T>^{J/\psi}_{NN}$,  the  mean  squared  transverse
momentum  in  NN  collisions and (ii) the
product of the gluon-nucleon cross-section and the average parton
momentum broadening per collision,  $\sigma_{gN}\delta_0$.  
$<p^2_T>^{J/\psi}_{NN}$ is  measured in RHIC energy
p+p collisions, $<p^2_T>^{J/\psi}_{NN} = 4.2 \pm 0.7$ $GeV^{2}$. As gluons are not free, the other parameter, $\sigma_{gN}\delta_0$ is
essentially non-measurable. Its value can be obtained from experimental data on  $p_T$ broadening 
of $J/\psi$. In SPS energy S+U/Pb+Pb collisions $\sigma_{gN}\delta_0$ is estimated as $0.442 \pm 0.056$ $GeV^2$ \cite{Chaudhuri:2006zg}. $\sigma_{gN}\delta_0$ at RHIC energy
is of interest.

PHENIX data on the centrality dependence of mean square transverse momentum $<p^2_T>$, in Cu+Cu and in Au+Au collisions are shown Fig,\ref{F5}.  
$<p^2_T>$ in
in p+p and in d+Au collisions are also shown. Quality of data is poor, few data points with large error bars.  Evidently, data do not show any  evidence of
$p_T$ broadening. Within the errors, $<p^2_T>$ in Cu+Cu  and in Au+Au collisions agree with that in NN collisions. $p_T$ broadening  of $J/\psi$'s is minimum at RHIC. 

To find $\sigma_{gN}\delta_0$ at RHIC energy, we fit the combined Cu+Cu and Au+Au data set (individual Cu+Cu or Au+Au
data points are few).  
We fix  $<p_T^2>_{NN}$ at the measured central value,
$<p^2_T>_{NN}$ = 4.2 $GeV^2$, and vary $\sigma_{gN}\delta_0$.
$<p^2_T>$ in Au+Au or in Cu+Cu show very little dependence
on the threshold density. In Fig.\ref{F5},
best fit is obtained  with threshold density
$n_c$=3.6,3.7,3.8,3.9 and 4.0 $fm^{-2}$ are shown. They
can not be distinguished.  
Best fit is obtained with  $\sigma_{gN}\delta_0 =0.31 \pm 0.48$ $GeV^2$. 
Due to poor quality of the data, the $\sigma_{gN}\delta_0$ is ill 
determined. Estimated error is larger
than the central value. We conclude that PHENIX data can not
determine the $\sigma_{gN}\delta_0$ at RHIC energy.

\section{summary and conclusions}

To conclude, in the QGP motivated threshold model, we have analyzed the preliminary PHENIX data
on the centrality dependence of $J/\psi$ suppression in
Cu+Cu and in Au+Au collisions. In the threshold model, in addition to the normal nuclear absorption, $J/\psi$'s are   anomalously suppressed, such that, if the local transverse density exceeds a threshold density $n_c$,
all the $J/\psi$'s are absorped. The model predicts a sudden change
of slope in $J/\psi$ suppression as a function of centrality.
Preliminary PHENIX data on the centrality (participant number) dependence of $J/\psi$ suppression in Au+Au collisions at RHIC are well explained in the model. It also
reproduces the sudden change of slope around participant number, $N_{part}$= 150.  Estimated threshold density ranges between 3.6-3.7 $fm^{-2}$. PHENIX preliminary data on the centrality dependence of $J/\psi$ suppression in Cu+Cu collisions do not show any sudden change 
of slope, characteristic of the threshold model. Local transverse 
density in most central Cu+Cu collision is only $2.63 fm^{-2}$,
much less than the estimated threshold density in Au+Au collisions. Anomalous suppression is not effective in Cu+Cu   
 collisions. However, $J/\psi$ production in Cu+Cu collisions is not explained the Glauber model of normal 
nuclear absorption.  While very peripheral collisions require $\sigma_{abs} \sim$1-3 mb, more central
collisions require $\sigma_{abs} \sim$6-7 mb. 
It appears that $J/\psi$ production in Cu+Cu collisions is
more complex than envisaged in the simple Glauber model of
nuclear absorption. 
We have also analyzed the
PHENIX data on $p_T$ broadening. The quality of data is not good enough for a definitive conclusion. Apparently at RHIC energy,  $J/\psi$'s donot show any $p_T$ broadening.

In conclusion, present analysis   strongly
support deconfined matter formation in central Au+Au collisions
at RHIC energy $\sqrt{s}_{NN}$=200 GeV.

 
\end{document}